# A Data-driven Spectral Model of Main-sequence Stars in Gaia DR3

Isabel Angelo[1], Megan Bedell[2], Erik Petigura[1], and Melissa Ness[2,3,4]
[1] Department of Physics & Astronomy, University of California Los Angeles, Los Angeles, CA 90095, USA
[2] Center for Computational Astrophysics, Flatiron Institute, 162 5th Avenue, New York, NY 10010, USA
[3] Research School of Astronomy & Astrophysics, Australian National University, Canberra, ACT 2611, Australia
[4] Department of Astronomy, Columbia University, 550 West 120th Street, New York, NY 10027, USA

## Abstract

Precise spectroscopic classification of planet hosts is an important tool of exoplanet research at both the population and individual system level. In the era of large-scale surveys, data-driven methods offer an efficient approach to spectroscopic classification that leverages the fact that a subset of stars in any given survey has stellar properties that are known with high fidelity. Here, we use The Cannon, a data-driven framework for modeling stellar spectra, to train a generative model of spectra from the Gaia Data Release 3 Radial Velocity Spectrometer (RVS). Our model derives stellar labels with precisions of 72 K in $T_{\rm eff}$, 0.09 dex in $\log g$, 0.06 dex in [Fe/H], 0.05 dex in [$\alpha$/Fe], and 1.9 km s$^{-1}$ in $v_{\rm broad}$ for main-sequence stars observed by Gaia DR3 by transferring GALAH labels, and is publicly available at https://github.com/isabelangelo/gaiaspec. We validate our model performance on planet hosts with available Gaia RVS spectra at SNR>50 by showing that our model is able to recover stellar parameters at ⩾20% improved accuracy over the existing Gaia stellar parameter catalogs, measured by the agreement with high-fidelity labels from the Spectroscopic Observations of Cool Stars survey. We also provide metrics to test for stellar activity, binarity, and reliability of our model outputs and provide instructions for interpreting these metrics. Finally, we publish updated stellar labels and metrics that flag suspected binaries and active stars for Kepler Input Catalog objects with published Gaia RVS spectra.

## 1. Introduction

Precise spectroscopic classification of stars has implications spanning all subfields of astronomy at both the population and individual system levels. Surveys like APOGEE (Majewski et al. 2017) and LAMOST (Cui et al. 2012) have facilitated great strides in galactic archeology and enabled the discovery of binaries, transients, and compact objects in the Milky Way. Spectroscopic observations of Kepler planet hosts illuminated features like the Radius Valley and Hot Neptune Desert that encode information about planet formation (e.g., Fulton & Petigura 2018; Petigura 2020). Beyond this, spectroscopic observations of individual systems provide details of the system's age, composition, and stellar properties (see, for example, Johnson et al. 2017; Rice & Brewer 2020).

Historically, spectroscopic classification has taken both data-driven and synthetic modeling approaches. The data-driven approach dates far back to the early 20th century, when Annie Jump Cannon classified thousands of stars by eye based on their optical spectra. In the following decades, spectroscopic classification shifted to rely on matching data with synthetic spectral models. However, these models are not always calibrated to real stars and are based on incomplete physics. For example, many of these models assume local thermodynamic equilibrium (LTE) or employ incomplete atomic line lists, simplified convection treatment, or nonphysical boundary conditions. More modern studies match data to real spectra from a library of well-characterized stars (e.g., Yee et al. 2017), and while these library spectra tend to resemble the data better than synthetic models, their performance decreases quickly for spectral resolutions below $R \sim 20{,}000$ (Behmard et al. 2019).

In recent years, data-driven spectral models have been used to infer stellar characteristics from spectra. These models are trained on large sets of spectroscopic data and predetermined stellar labels like effective temperature, surface gravity, and elemental abundances using supervised learning algorithms to produce generative models of stellar spectra (see, for example, The Cannon; Ness et al. 2015 and the Data-Driven Payne; Xiang et al. 2019). These models have been employed for numerous science cases, from mapping elemental abundances (e.g., Ness et al. 2016; Buder et al. 2018) to identifying binaries (El-Badry et al. 2018b) and constraining properties of planet hosts (Behmard et al. 2019; Rice & Brewer 2020). One of the key features of data-driven spectral models is that they leverage the fact that for a given survey, a subset of stars often have derived labels with higher fidelity from a previous survey. In training a model on these predetermined labels, along with spectra from the survey of interest, precise and accurate spectral information encoded in the model is transferred across data sets. The model can then be used to derive high-fidelity labels for the rest of the stars in the survey of interest.

Nearly 1 million near-infrared spectra are included in Gaia Data Release 3 (DR3), and this number is expected to increase by ∼2 orders of magnitude in future data releases (Katz et al. 2023). These spectra cover a wavelength range of 845–872 nm with a spectral resolution of $\lambda/\Delta\lambda \sim 11{,}500$, with spectral information in the Ca II IR triplet at wavelengths of 849.8 nm, 854.2 nm, and 866.2 nm, as well as shallower lines from various other elements. While the large sample size lends itself well to large-scale classification and demographic analysis, the spectra have smaller bandwidth and lower signal-to-noise ratio (SNR) than a dedicated ground-based survey can typically achieve (e.g., GALAH, APOGEE, etc., see De Silva et al. 2015; Majewski et al. 2017). A data-driven spectral model trained on higher fidelity labels from one of these surveys can transfer pre-existing spectral information onto this new data set and assess how far we can extrapolate high-fidelity labels to the growing sample of Gaia Radial Velocity Spectrometer (RVS) spectra. For example, this method was shown to be effective for retrieving precise refractory elemental abundances of solar analogs in Gaia Rampalli et al. (2024), see.

In this paper, we use The Cannon's data-driven framework to train a spectroscopic model of the Gaia RVS spectra using a set of well-characterized main-sequence stars from GALAH. We outline the methods that The Cannon uses to train a model and use it to retrieve stellar labels in Section 2 and describe the data we train our model on in Section 3. In Section 4, we update our implementation of The Cannon and validate our model performance. In Section 5, we evaluate our model performance by measuring its agreement with high-fidelity labels from the Spectroscopic Observations of Cool Stars (SPOCS) survey, and compare it to the performance of existing Gaia catalogs. We also publish updated stellar labels for stars in the Kepler Input Catalog (KIC) with published RVS spectra from Gaia DR3. We discuss applications for identifying unresolved binaries in the Gaia RVS spectra in Section 6.2 and outline metrics that our model reports for identifying active stars, evolved stars, and cases where our model performance is compromised in Section 6. We summarize our results in Section 7. Our model (Angelo 2024) is available at https://github.com/isabelangelo/gaiaspec to enable the characterization of user-specified stars with RVS spectra from DR3 or future data releases.

## 2. The Cannon Overview

We used The Cannon 2, an open-source implementation of The Cannon introduced in Casey et al. (2016), to train our data-driven model. In this section, we outline the basic methods of this implementation to train a Cannon model and fit a trained model to new spectra to derive stellar labels. A more detailed description of these methods can be found in Ness et al. (2015) and Casey et al. (2016).

The Cannon interacts with spectroscopic data in two ways: it uses spectra during its "training step" to develop a spectral model and then uses this model to fit to individual spectra during its "test step." To summarize, the training step determines the relationship between the observed flux and predetermined "ground truth" labels at each individual pixel for all the stars in the training set. The test step fits a trained Cannon model to the observed flux of a star with labels to be determined by The Cannon model itself. These steps are described in detail below.

### 2.1. Training Step: Supervised Learning of "Ground Truth" Spectra

The Cannon is a generative linear regression model that predicts the flux of an object as a function of the labels it's trained on (in our case, $T_{\rm eff}$, $\log g$, [Fe/H], [$\alpha$/Fe], $v_{\rm broad}$). The $v_{\rm broad}$ parameter here is a line-broadening parameter that is determined by the combined effect of stellar rotation and macroturbulence, and is derived for both GALAH and Gaia RVS spectra (Buder et al. 2021; Frémat et al. 2023). A description of The Cannon can be found in Ness et al. (2015) and Casey et al. (2016).

During the training step, The Cannon uses a supervised learning algorithm that takes stellar labels and flux as input. The goal during this step is for the model to learn how the flux at a particular wavelength interval, or "pixel" depends on a star's labels. The training step learns from The Cannon model using the underlying assumptions that (1) stars with the same labels have the same per-pixel flux, (2) the flux at a particular pixel is a polynomial function of the training labels, and (3) the flux at a particular pixel is independent of the flux at neighboring pixels. Note that the third assumption means that the pixel fits are performed independently of one another–the model does not assume any "smooth" behavior across pixels. The labels are treated as ground truth, and for each pixel, the training step fits a polynomial to the relationship between the flux and the stellar labels.

For a particular pixel value $j$ and stellar $n$, The Cannon assumes that the flux $f_{n,j}$ depends on the star labels $l_n$ according to the following relation:

$$f_{n,j} = v(l_n) \cdot \theta_j + e_{n,j}. \qquad (1)$$

In this case, the "vectorizer" $v(l_n)$ is a function that describes how the flux is assumed to vary with the labels. For our model, we assume that the pixel flux varies linearly to second order with the five stellar labels so the vectorizer contains 21 terms (one linear and one quadratic term for each label, with 10 cross terms and one constant). The coefficients of the vectorizer polynomial are contained in $\theta_j$, and $e_{nj}$ are the flux errors, which for this paper are sampled from a Gaussian with zero mean and variance equal to the instrumental uncertainty and intrinsic pixel scatter of the model summed in quadrature, $\sigma_{nj}^2 + s_j^2$. We note that while The Cannon 2 computes a model-specific $s_j^2$ during its training step, we derive an empirical $s_j^2$ that we use throughout this analysis (see Section 4.2).

In the training step, the flux ($f_{nj}$), flux errors from photon noise and instrumental effects ($\sigma_{nj}$) and the labels of the spectra ($l_n$) are known, but the $\theta_j$ coefficients are not. The $\theta_j$ coefficients govern the model behavior by predicting the flux at a given pixel according to Equation (1). The log-likelihood of the model flux at a particular pixel is:

$$\ln p(f_{nj}|\theta_j, l_n, s_j^2) = -\frac{[f_{nj} - v(l_n) \cdot \theta_j]^2}{s_j^2 + \sigma_{nj}^2} - \ln(s_j^2 + \sigma_{nj}^2). \qquad (2)$$

The training step determines the $\theta_j$ coefficients and $s_j^2$ associated with each pixel that maximize this likelihood function (or rather, minimizes the negative likelihood function) for all reference stars $N_{\rm ref}$ in the training set:

$$\theta_j, s_j^2 = \arg\min_{\theta_j, s_j^2} \sum_{n=1}^{N_{\rm stars}} -\ln p(y_{nj}|\theta_j, l_n, s_j^2), \qquad (3)$$

thereby computing the $\theta_j$ coefficients that best describe the training flux. Once it has computed these best-fit coefficients, it can predict the flux at each pixel for a given set of stellar labels based on Equation (1) using the $\theta_j$ that is derived in the training step. This is described further in Section 2.2.

We note that due to the data-driven nature of The Cannon, our model performance is limited by what we will refer to hereafter as the training domain. The training domain is a 5D space with each dimension corresponding to a single training label, populated by data points that each represent a single spectrum that the model was trained on. In the event that The Cannon identifies best-fitting labels beyond this domain, the best-fit spectrum was achieved via extrapolation beyond the label range of the reference objects used to build the main-sequence model. In this case, the labels are not tied to any ground truth (see Section 4.3). Additionally, The Cannon's label retrieval can only be as accurate and precise as the "ground truth" labels it's trained on.

### 2.2. Test Step: Model Fitting to Derive Stellar Labels

Once The Cannon model is trained, we can use it to recover labels of stars outside the training set in the test step, which fits a trained Cannon model to spectroscopic observations of the star to be characterized. In the training step, the flux and labels of the training set are known, and the optimizer is fitting for $\theta_j$ and $s_j^2$. In the test step, the labels are unknown, but the values for $\theta_j$ and $s_j^2$ have been predetermined by the test step. Thus, for a given spectrum, the optimal label vector $\hat{l}_n$ can be described as the labels associated with the maximum-likelihood Cannon model flux:

$$\hat{l}_n = \arg\min_{l_n} \sum_{j=1}^{N_{\text{pixels}}} -\ln p(f_{n_j}|\theta_j, l_n, s_j^2) \qquad (4)$$

$$= \arg\min_{l_n} \sum_{j=1}^{N_{\text{pixels}}} \frac{[f_{nj} - v(l_n) \cdot \theta_j]^2}{s_j^2 + \sigma_{nj}^2} \qquad (5)$$

where $-\ln p(f_{n_j}|\theta_j, l_n, s_j^2)$ (hereafter $\chi^2$) represents the goodness of fit for a model with label vector $l_n$. Although The Cannon 2 implementation that we used offers its own test step implementation,[5] we wrote our own custom function that allows us to place further constraints on our model extrapolation during the test step, as described further in Section 4. We confirmed that our implementation performs just as well on ground truth label retrieval as The Cannon 2's implementation. An illustration of the coefficient vectors that The Cannon computed during our model's training step can be found in the Appendix.

### 3. Training Data

The Cannon requires a set of well-characterized stars with rest-frame, continuum-normalized spectra on a uniform wavelength scale. It also requires the corresponding labels for all of those stars. For our model, we train on labels $T_{\text{eff}}$, $\log g$, [Fe/H], [$\alpha$/H], $v_{\text{broad}}$. The stars and their labels should span the domain in which you want your model to perform well. Because The Cannon is measuring the flux dependence on the labels at the individual pixel level, spectra with high SNR and accurate, precise labels is best.

One million spectra of "well-behaved"[6] objects are available as part of the Gaia Data Release 3 (DR3; Cropper et al. 2018). These spectra have spectral resolution $R = \Delta\lambda/\lambda = 11,500$ over the infrared range $\lambda = 845$–$872$ nm. The spectra are rest-frame shifted, continuum normalized, and averaged over multiple visits over 34 months with an average of eight transits per star per year (Katz et al. 2023). This, along with the large number of available spectra, makes them ideal for implementing a Cannon model. However, in order for our model to perform with higher precision and accuracy than the published Gaia stellar labels, we need to use higher fidelity reported labels for the stars in our training set.

For the labels corresponding to stars in our training set, we used labels computed by the GALAH+ survey's third data release (Buder et al. 2021). This survey published derived stellar labels for 588,571 nearby objects, including a large number of main-sequence stars that were relevant as potential training objects for our model. Their labels were derived using a combination of GALAH spectra and Gaia DR2 and Two Micron All Sky Survey photometry to improve the accuracy of the derived labels beyond the accuracy that the spectra alone can provide. Beyond this, GALAH spectra have higher resolution and more wavelength coverage than Gaia ($R = 28,000$ over four 25 nm channels between 470 and 790 nm), thereby providing higher fidelity labels that are ideal for label transfer. Their catalog also included a cross match with Gaia DR3, which allowed us to map derived labels to the spectra in our training set.

Our original sample contained 588,571 objects with GALAH-reported labels. From here, we applied a number of cuts to obtain a well-characterized sample of single stars that would optimize our model's performance. The cuts we made can be subdivided into three groups based on their broader purpose, as follows:

1. *Quality of spectral data.* We required an SNR of at least 100/pixel for both the GALAH reference object (catalog label `snr_c3_iraf`) and Gaia RVS spectrum (catalog label `rvs_spec_sig_to_noise`) for each object in our training set. We also removed objects from GALAH that did not have a published Gaia RVS spectrum.[7]

2. *Confidence in stellar labels.* We excluded objects that were flagged for poor label quality in the GALAH catalog. To enforce this, we required the stellar label, iron abundance, and $\alpha$ abundance quality flags (`flag_sp`, `flag_fe_h`, and `flag_alpha_fe`, respectively) to be zero. We also removed stars with any label whose uncertainty was more than twice the median uncertainty of the corresponding label in the full GALAH catalog. Finally, we required finite GALAH-reported values for all five of our training labels.

3. *Contamination from unresolved binaries.* In the event that the stars in our training sample are unresolved binaries, the GALAH-reported stellar labels will be biased, and the model performance is compromised (El-Badry et al. 2018b, hereafter EB18). To ensure that the stars in our training set are all single stars, we removed GALAH stars that were identified as binaries in Traven et al. (2020). We also required stars to be well fit by single-star astrometric solutions. This information is encoded in the object's Renormalised Unit Weight Error (RUWE) value and Gaia `non_single_star` value (see Belokurov et al. 2020). For our training set, we require Gaia DR3 and Gaia DR2 RUWE < 1.4 and Gaia `non_single_star`=0. Finally, we remove potential binaries missed by the Gaia and GALAH detection methods with an iterative process described in Section 4.4.

---

[5] https://github.com/andycasey/AnniesLasso
[6] stars with SNR⩾15, $3100 < T_{\text{eff}} < 14{,}500$ K, and no problems introduced by the processing pipeline (Katz et al. 2023).
[7] The SNR>100 threshold was chosen because it was the lowest SNR threshold for which we were able to achieve output label precision comparable to the GALAH-reported label errors for stars in our training set.

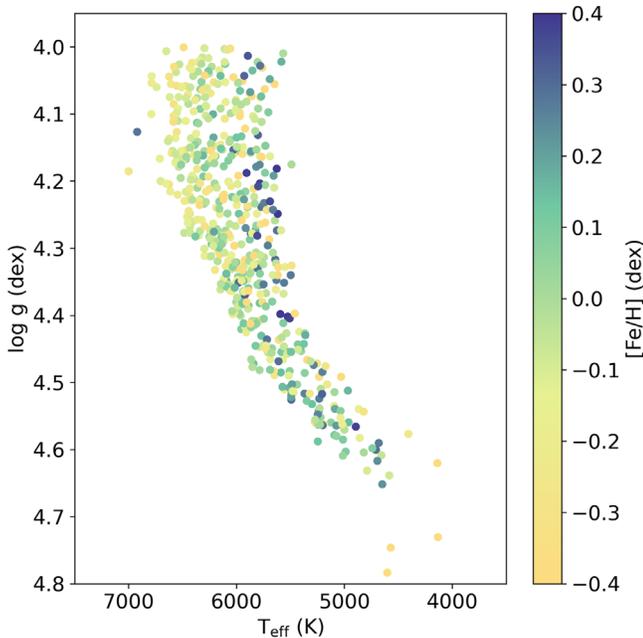

**Figure 1.** Distribution of $T_{eff}$, $\log g$, and [Fe/H] values for the stars in our model training set. The points represent individual stars in a 3D subset of our 5D training domain. Our model training set contains 563 well-characterized main-sequence stars from GALAH with high-fidelity labels.

We also limited our training set to main-sequence stars by requiring $\log g > 4$. Of the 588,571 stars in the original GALAH sample, 28,530 remained after removing stars based on GALAH-reported SNR or quality flags, and 8111 remained after removing stars with large label uncertainties, $\log g < 4$ and binaries identified by Traven et al. 2020. Only 632 of these had published RVS spectra with SNR >100, and this number was reduced to a final training set size of 563 after we iteratively removed potential binaries (see Section 4.4 for a description of this process). Figure 1 shows the $T_{eff}$, $\log g$, [Fe/H] domain of our final training set.

Due to the fact that the RVS spectra are rest-frame shifted, the RVS spectra available in Gaia DR3 often have pixels with NaN values at either the beginning or the end of the spectrum. For our training set, ~10% of the spectra had NaN values in the first or last 20 pixels of the spectra. To address this, we clipped all of the spectra by 20 pixels on each side before training The Cannon. After clipping, only ~1% of training spectra had NaN pixels remaining in the spectra. To preserve the information in the 99% of spectra with finite values for these pixels, we interpolated the flux at these values and inflated their reported errors to infinity to effectively mask them during the training step.

## 4. Model Customizations

Our trained Cannon model predicts the flux at each RVS pixel as a function of the training labels over a wide range of parameter space (the trained model coefficients for each label can be found in the Appendix). From there, we made a number of changes to the original Cannon test step described in Section 2.2 to improve our model performance and conducted tests to ensure that we could trust our model's output labels. The key customizations are described in the remainder of this section. We confirmed that our implementation performs just as well on ground truth label retrieval as The Cannon 2's implementation.

### 4.1. Masking the Calcium IR Triplet

The Gaia RVS spectral range includes three deep calcium lines at 849.8 nm, 854.2 nm, and 866.2 nm, collectively known as the Ca-triplet. Certain astrophysical phenomena can manifest as superimposed emission and absorption confined to only these features in the RVS spectrum. Examples include interacting binaries, young accreting T Tauri stars, chromospheric activity in GKM main-sequence stars, emission linked to the accretion of matter in the stellar magnetosphere, etc. (see Lanzafame et al. 2023, for a description). While we did not explicitly remove active stars from our training set, we assume that since our training set is composed primarily of inactive, main-sequence stars, the effects of activity would not be significant enough for The Cannon to learn for pixels outside the Ca-triplet. In other words, The Cannon's assumption that flux depends solely on static labels breaks down at the Ca-triplet, where time-dependent surface features and activity contribute to the flux as well. This was demonstrated by Rampalli et al. (2021), who trained a polynomial regression model on RVS spectra and found that it over- or underpredicted flux at the Ca-triplet for 82% of the stars in their sample.

Given this assumption, while our best-fit Cannon model should produce a good fit to the spectrum at all the lines when the star is inactive, this should not be the case for active stars. Instead, for stars with activity-induced calcium emission, The Cannon model should be a good fit everywhere except the calcium features. In these cases, our Cannon model might fit to these features at the cost of providing a good fit to the rest of the spectrum. Thus, our model performance is compromised when we include these features in our Cannon test step. To avoid this, we mask out the Ca-triplet when we perform a model fit to a spectrum by setting their flux errors equal to be infinite. We note, however, that this mask is not driven by a need in the data; because stellar activity affects flux in the cores of the Ca-triplet lines and is not included as a training label, we infer from this knowledge that the model performance is compromised for these features.

In the published model and validation steps in Section 4.2, we are fitting to the nonshaded parts of Figure 2. We stress that since The Cannon fits each wavelength pixel independently, including these masked pixels in the training set does not introduce any bias in the rest of the spectrum. Consequently, our trained Cannon model still predicts the flux for the Ca-triplet features as a function of stellar labels, though we caution that these predictions may not be reliable.

Figure 2 shows a test case where our calcium mask is essential for recovering the correct labels for the spectrum. We show an active dwarf identified by Lanzafame et al. (2023) with a Cannon model fit that was computed with a mask on the calcium features. We see that the model is a good fit to the entire spectrum, except for the calcium features, which are shallower than the model predicts due to emission at those features that our model does not account for. Additionally, the best-fit model for a masked spectrum predicts what the flux at the Ca-triplet should be for that star in the absence of activity. Thus, we can identify active stars in cases where the true flux deviates significantly from the flux of the best-fit model computed with the Ca-triplet mask (see Section 6.1 for a detailed description of this process).

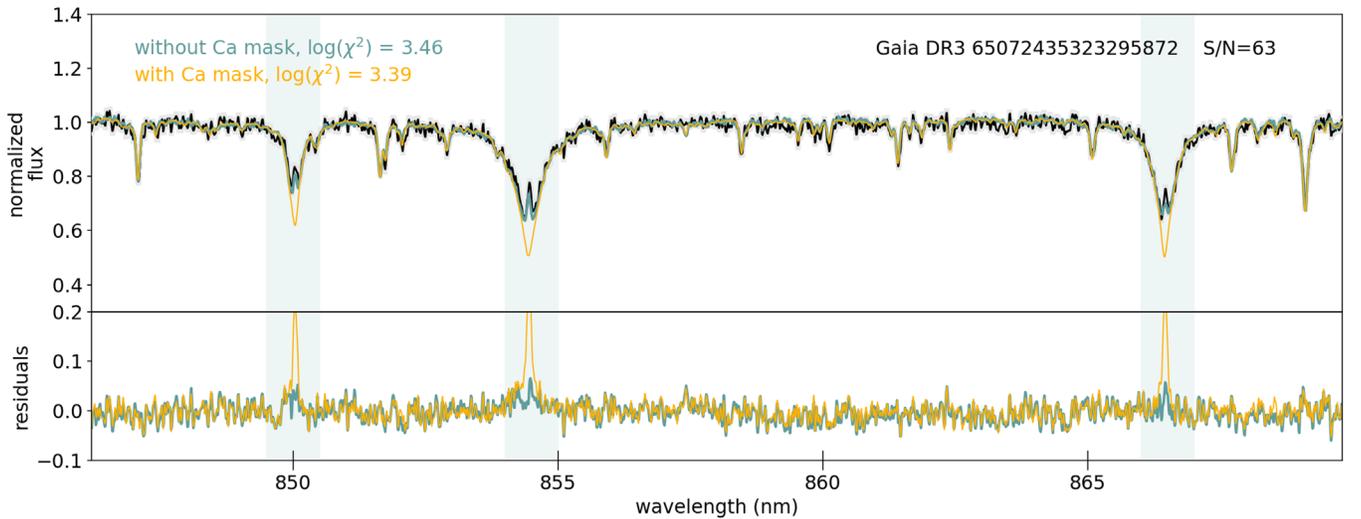

**Figure 2.** Demonstration of why a calcium mask is needed to correctly derive labels for RVS spectra. The spectrum shown in black is an active dwarf star with superimposed emission and absorption in the Ca-triplet lines. Without masking the shaded regions, The Cannon fits to the Ca-triplet at the expense of finding a good fit to the other features (shown in blue). When we include a mask, the best-fit model describes the spectrum well everywhere except the calcium lines (shown in yellow).

### 4.2. Empirical Model Scatter

As outlined in Section 2, our model uses a quantity called $s^2$ to compute the $\chi^2$ goodness of fit associated with a particular set of labels for a given spectrum. This term describes the per-pixel intrinsic model scatter, or in other words, it encapsulates any pixel-level variation in the training set spectra that cannot be attributed to photon noise. When we first trained our model, the training step found $s^2$ to be $\sim 10^{-4}$ at the 3 pixels corresponding to the centers of the calcium line cores and zero for all other pixels (see Casey et al. 2016, for a description of how the training step computes $s^2$). While this original $s^2$ is able to compute fits with reasonable $\chi^2$ for SNR$\sim$50–100 where the deviation from the model is dominated by photon noise and instrumental effects, fitting our model to higher SNR targets from the SPOCS and CKS samples revealed a clear correlation between SNR and $\chi^2$ for stars with SNR $\gtrsim$200. This is because stars with significantly higher SNR than the training set possess intrinsic scatter that is lost to photon noise for the SNR$\sim$100 training set stars. In other words, the $s^2$ originally computed in our Cannon training step was under-estimated, inflating the $\chi^2$ for targets with high SNR.

To address this, we computed an empirical per-pixel model scatter that we adopt for our models $s^2$ term. We gathered all stars in Gaia with published SNR $\geqslant$500 RVS spectra and labels that reside reasonably within the training domain (log $g$>4 dex, 4000< $T_{\rm eff}$ <8000 K). We removed obvious binaries by requiring RUWE <1.4, `non_single_star`=0 and downloaded the 94 remaining RVS spectra. We expect the primary deviation from our model to be due to intrinsic model scatter for these spectra since their photon noise is so low. Thus, for each model we computed the best-fit Cannon model and recorded the empirical scatter between the true flux and the best-fit model (i.e., |data − model|$^2$). We repeated this calculation for each of the spectra in our high-SNR sample and computed the average across all stars for each pixel to arrive at an empirical $s^2$ prediction. We found that the best-fit Cannon model differs from the RVS spectra by $s^2 = 0.1\%$–1% for these objects, which is more than the $\leqslant$0.1% variations predicted from their photon noise alone. We verified that including our empirical $s^2$ term successfully removed the observed correlation between high SNR and $\chi^2$ that we saw for the SPOCS sample originally. Thus, we use the empirical $s^2$ term for our model throughout the entire analysis of this paper. A plot of our per-pixel empirical $s^2$ can be found in the Appendix.

We do not expect the training step to infer zero scatter for a training set with accurate flux errors. However, if the flux errors are overestimated, they may be sufficient to explain the model's deviation from the true fluxes without any intrinsic model scatter. Indeed, when we computed best-fit Cannon models for a sample of main-sequence single stars in Section 6, we found that the distribution of $\chi^2$ values peaks at $\sim$2150, which is slightly smaller than the expected peak near the total number of pixels in the spectrum (2361 for our training set). This indicates the Gaia RVS flux errors may be overestimated for the Gaia RVS spectra. Thus, we suspect that overestimated flux errors are the reason that our model's test step computes a zero scatter during its optimization.

### 4.3. Constraints on Extrapolating beyond the Training Domain

During the test step outlined in Section 2, The Cannon is free to extrapolate to labels outside of the training domain. In this regime, however, the labels that best fit the model spectrum are not tied to any ground truth, and the labels output by the model are less reliable. To address this, we place constraints on the degree to which we allow The Cannon to extrapolate beyond the training domain during the test step. We measure the extrapolation using a 5D Gaussian kernel density estimation (KDE; Rosenblatt 1956), which uses nonparametric kernel smoothing to approximate the probability density of stars in the training domain. We refer to this quantity hereafter as the training label density, $\rho(l_n)$. The training label density is a function of the five Cannon model training labels ($T_{\rm eff}$, log$g$, [Fe/H], [α/H], $v_{\rm broad}$), and the output represents the density of training set objects near those labels in the training domain. A Cannon fit that returns labels with a high $\rho(l_n)$ is the product of a reasonable interpolation within the training domain, thereby producing a reasonable model fit with accurate labels. In contrast, a Cannon model with a low $\rho(l_n)$ is extrapolating beyond the training domain to produce a nonphysical model with unreliable labels and an artificially inflated goodness of fit.

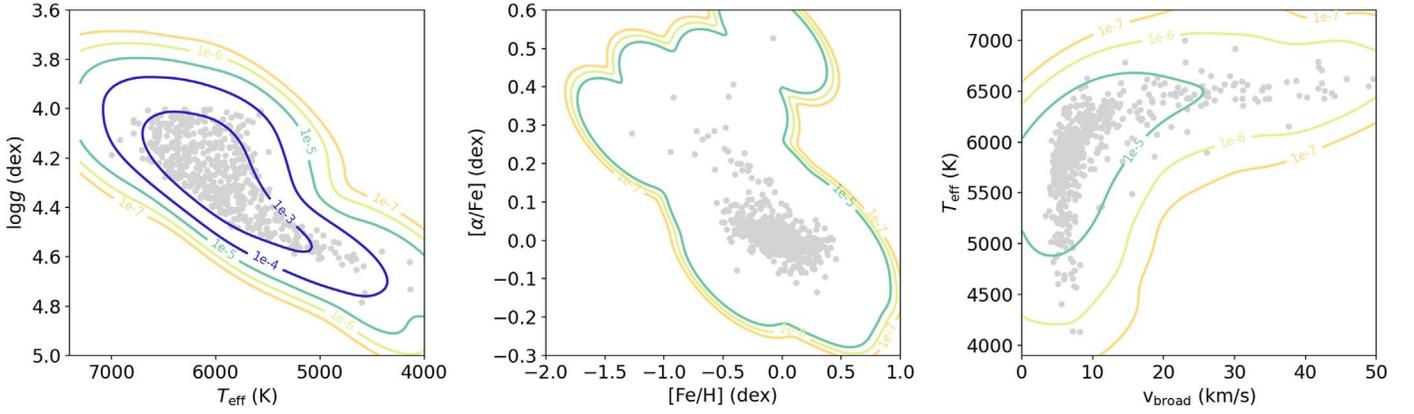

**Figure 3.** Illustration of our model's $\rho(l_n)$ parameter. We plot the distribution in 2D, where each point represents a model in our training density. Densely populated regions of this space correspond to domains with high $\rho(l_n)$. We overplot lines of constant $\rho(l_n)$ in each panel. Our model penalizes labels that lie beyond the orange lines, where $\log \rho(l_n) > -7$.

To prevent The Cannon from extrapolating beyond the training domain when predicting labels, we determined a reasonable $\rho(l_n)$ range to allow The Cannon to model by inspecting the $\rho(l_n)$ distribution for all of the stars in our training set. Figure 3 shows all of the stars in the training set as a function of their labels, with contours to illustrate lines of constant KDE-estimated $\rho(l_n)$. We used a leave-one-out technique where for each star, the KDE used to determine $\rho(l_n)$ is calculated using the full training set with the individual star held out. This prevents the distribution from piling up around $\log \rho(l_n) = -4$ due to the fact that the object we were evaluating went into estimating the KDE that was used to compute its $\rho(l_n)$. During our test step, we calculate $\rho(l_n)$ using a KDE computed with every object in the training set.

As shown in Figure 3, objects in the training set typically reside in densely populated regions of the training domain, with $\rho(l_n) \sim 10^{-3}$. This number falls off quickly, and very few stars in the training set have $\rho(l_n) \sim 10^{-7}$. Thus, to prevent our model from extrapolating beyond the training domain, we updated the $\chi^2$ value that is minimized in the test step (Equation (4)) to include an inflation factor that penalizes models with $\rho(l_n) < 10^{-7}$:

$$\hat{l}_n = \arg\min_{l_n} \sum_{j=1}^{N_{\rm pixels}} -\ln p \times P(l_n) \qquad (6)$$

where $P(l_n)$ is a penalty factor that penalizes sets of labels $l_n$ with $\rho(l_n) < 10^{-7}$:

$$P(l_n) = \begin{cases} 1, & \text{if } \rho(l_n) > 10^{-7} \\ 1 + \log\left[\dfrac{10^{-7}}{\rho(l_n)}\right], & \text{if } \rho(l_n) \leqslant 10^{-7} \end{cases}. \qquad (7)$$

We do not include a penalty term for $\rho(l_n) > 10^{-7}$ to avoid favoring models simply for being in the most populated regions of training domain. The addition of 1 to the fractional penalty ensures that the likelihood function is continuous at $\rho(l_n) = 10^{-7}$. We also sample $\log v_{\rm broad}$ instead of $v_{\rm broad}$ in our optimization routine to avoid fitting with nonphysical $v_{\rm broad} < 0$ values.

### 4.4. Removing Spectral Binaries from the Training Set

While the binary cuts outlined in Section 3 remove binaries that were previously identified by Gaia, it is possible that our original training set is contaminated by unresolved spectroscopic binaries. In particular, binaries with separations wide enough to produce overlapping spectral features are often mistaken for single stars and are likely abundant in the RVS sample. In the event that our training set is contaminated with these "spectral" binaries, the training step will not accurately learn how flux depends on labels for single stars, and our Cannon model is not reliable.

To address this, we follow a similar procedure to that outlined in EB18 and iteratively remove these binaries from our training set. In brief, we train an original Cannon model on our initial training set and use this model to identify binaries (see Section 6.2 for a detailed explanation of this process). We then remove the identified binaries from the training set and train the second model with the "cleaned" trained set, and repeat this process until the model no longer identifies binaries in its own training set. We adopt conservative criteria for binaries in this step and remove spectra with $\log \Delta\chi^2 \geqslant 2$ from the training set at each iteration, where $\Delta\chi^2 = \chi^2_{\rm single} - \chi^2_{\rm binary}$ describes how well the binary model fits the data compared to the single-star model, as described further in Section 6.2.

## 5. Model Performance

### 5.1. Validation against GALAH Labels

We investigated how well our model is able to predict stellar labels by running the test step on the spectra into the training set and compared The Cannon–inferred labels to known "ground truth" labels from GALAH. For this, we used leave-one-out cross validation, where for each object in the training set we trained a new model with a new training set that is identical to the original, but with the object of interest held out. We then used each model to run the test step on the object that was held out of its training set. This allowed us to derive labels for each star in the training set without the bias introduced when the test step is run on an object that was in the model's original training set. Figure 4 shows the GALAH labels plotted versus The Cannon output labels found using The Cannon test step outlined in Section 2.2. The diagonal line represents the one-to-one line, where points should fall if our model is in perfect agreement with the ground truth labels.

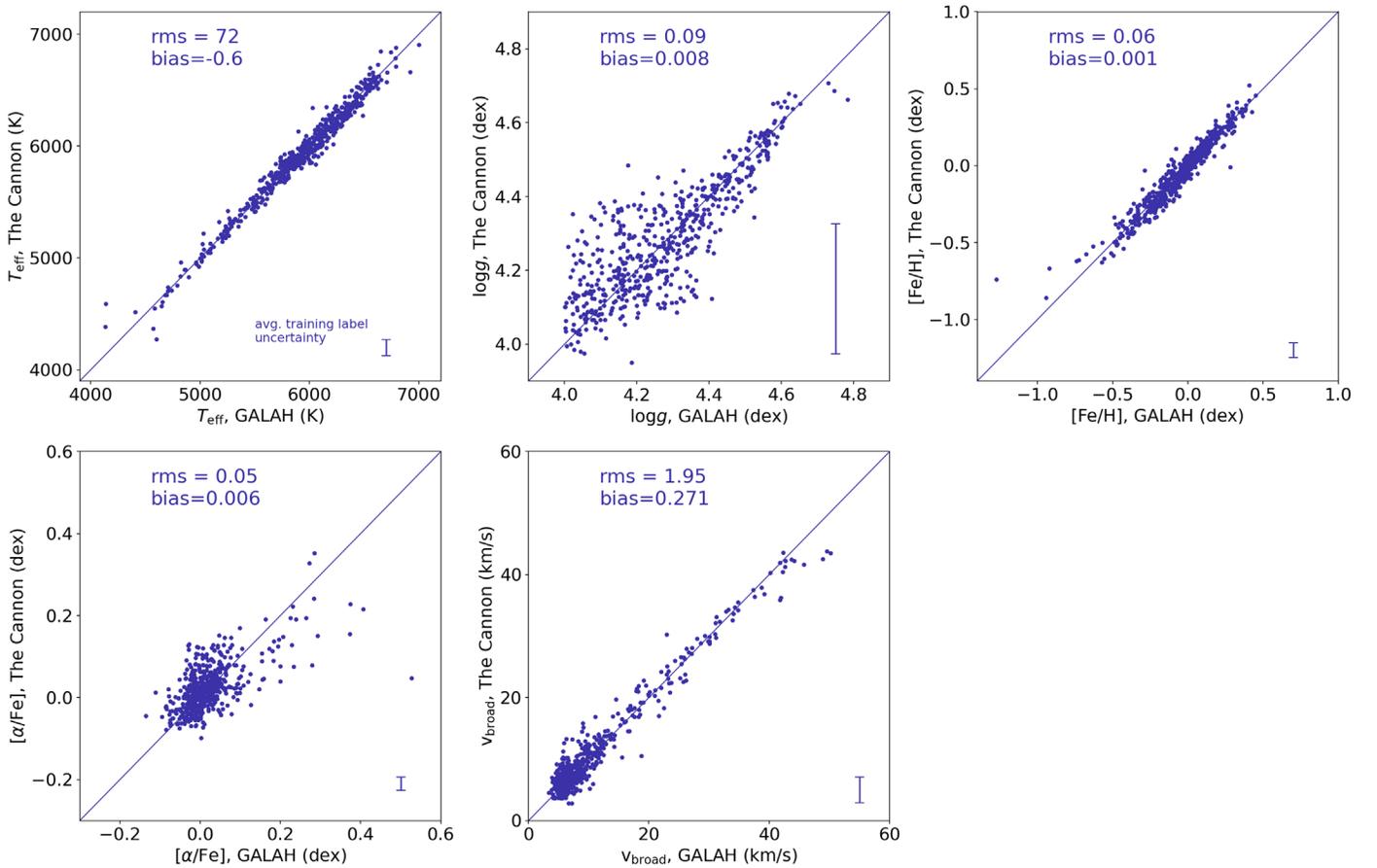

**Figure 4.** Cannon label outputs are plotted vs. "ground truth" labels from the GALAH survey for each star in our training set. The Cannon output labels are computed using the leave-one-out cross validation described in Section 4. The rms scatter and bias values are shown to quantify the average disagreement between The Cannon output and GALAH labels.

The rms difference between the GALAH and The Cannon labels is listed in Table 1, along with the range for each label seen by our training set.[8]

A Cannon model's performance is limited by the precision of its input training set labels, thus the model is performing well when its agreement with the training labels is comparable to the label uncertainties. We overlay the average training set label uncertainties in Figure 4. As we can see from the figure, the model uncertainties in Table 1 are comparable, albeit slightly larger, than the average GALAH-reported uncertainties. The difference can likely be attributed to the difference in data quality for the two surveys—the GALAH spectra have higher resolution ($R \sim 28{,}000$, compared to 11,500 for Gaia) and a longer bandpass ($\sim 300$ nm, compared to 25 nm for Gaia). Additionally, the RVS spectra of the objects in our training set are noisier than those in GALAH, with an average SNR $\sim 139$ and 168, respectively. The only exception to this is $\log g$, where our Cannon model agrees with the GALAH-reported $\log g$ to within 0.09 dex, which is less than the average uncertainty of 0.18 dex for our training set. This is likely due to the fact that The Cannon learns only the statistical uncertainty in training set labels as opposed to any systematic uncertainties that may be present in the GALAH labels. Thus, if the GALAH labels have systematic errors that affect multiple measurements of the same star in the same way, the GALAH error bars will reflect this, but The Cannon/GALAH rms scatter we report will not.

---
[8] While these ranges represent a useful 1D approximation of whether a given set of labels is the training set domain, $\rho(l_n)$ provides a better estimation.

**Table 1**
Average Cannon Model Disagreement with "Ground Truth" GALAH Labels for Training Set Stars

| Label | rms (GALAH—Cannon) | Training Set Range |
|---|---|---|
| $T_{\rm eff}$ | 72 K | 4132–6999 K |
| $\log g$ | 0.09 dex | 4.00–4.78 dex |
| [Fe/H] | 0.06 dex | −1.27 to 0.45 dex |
| [α/Fe] | 0.05 dex | −0.13 to 0.53 dex |
| $v_{\rm broad}$ | 1.9 km s$^{-1}$ | 3.34–50.23 dex |

### 5.2. Performance on Planet Hosts

To investigate our model's ability to recover labels of planet hosts, we compared our Cannon output labels to labels reported by the Spectroscopic Properties of Cool Stars (SPOCS; Brewer et al. 2016). The SPOCS sample is a benchmark catalog of high-fidelity stellar properties inferred from high-resolution ($R \sim 50{,}000$) high-SNR ($\gtrsim 100$/pixel) spectra of 1615 main-sequence planet hosts using a bespoke line-by-line spectral analysis of high-resolution spectra from the Keck High Resolution Echelle Spectrometer (HIRES). Of the 1615 individual stars with SPOCS spectra and labels reported in Brewer et al. (2016), 107 are main-sequence stars with published RVS spectra and reported labels spanning our model's training domain.

We used our model to compute best-fit models and corresponding stellar labels for each of these Gaia RVS spectra. The right panel of Figure 5 shows how well our Cannon-inferred labels agree with those reported by SPOCS for all 107 stars, and Figure 5 shows the same for the 80-star subset with RVS SNR >50. As we can see from these figures, the degree to which our model agrees with the high-fidelity SPOCS labels is largely limited by the SNR of the RVS spectrum that ours fits to. For RVS spectra with SNR >50, we find that our model reproduces stellar labels that agree with SPOCS labels to within 85 K in $T_{\rm eff}$, 0.12 dex in $\log g$, 0.09 dex in [Fe/H], and 2.3 km s$^{-1}$ in $v_{\rm broad}$. These label precisions indicate a 40% improvement over the Gaia `GspSpec` catalog's ability to reproduce the SPOCS labels, and a 20% improvement over the `GspPhot` catalog.

The improved agreement with SPOCS labels over preexisting Gaia catalogs has promising implications for constraining labels of exoplanet hosts without dedicated ground-based follow-up. In particular, for the majority of Kepler planet hosts, the most updated labels come from the Kepler Input Catalog (KIC), which published labels for stars in the `Kepler` field derived from stellar broadband photometry (Brown et al. 2011). These labels have large uncertainties (200 K in $T_{\rm eff}$ and 0.4 dex in $\log g$ and [Fe/H]), thus the characterization of these planets is limited by the degree to which their host properties are known. We used the Gaia–Kepler cross-match database[9] to identify main-sequence KIC stars with published RVS spectra with SNR>50 and $T_{\rm eff}$ within our training domain. We used our model to compute updated labels for these stars, along with their $\chi^2$, $\rho(l_n)$, and $\chi^2_{\rm Ca}$ values (see Section 6 for a description of these metrics). Table 2 shows our reported labels for these stars.

### 5.3. Comparison to Gaia-reported Labels

Our Cannon model reports $T_{\rm eff}$, $\log g$, [Fe/H], [$\alpha$/H], $v_{\rm broad}$ for any of the million stars with RVS spectra. Gaia reports these labels as well. Additionally, the Gaia `GspPhot` catalog reports stellar labels derived from the Bp/Rp spectra, which exist for ∼100 million stars in Gaia DR3 (Andrae et al. 2023). Stellar labels derived from the RVS spectra are also reported in the `GspSpec` catalog using MatisseGauguin and Artificial Neural Net (ANN) algorithms (Recio-Blanco et al. 2023). To place our model in the context of the existing Gaia catalogs, we compared its output labels, the `GspPhot` labels, and the `GspSpec` labels to "ground truth" SPOCS labels for the same subsample of SPOCS stars from Section 5.3. We note that in the case of [Fe/H], SPOCS and Gaia `GspPhot` report [M/H] instead of [Fe/H], so the comparison may not be strictly one to one. However, since [Fe/H] and [M/H] are correlated and the model learns from every pixel and not just the iron lines, the [Fe/H] we predict should be an adequate predictor of M/H as well.

The labels from all three sources are plotted against the SPOCS labels in Figure 5, with the rms scatter and bias listed for each panel. The subset of stars with Gaia RVS SNR>50 is outlined to show the improvement in our model performance for high SNR. Our Cannon model agrees more closely with the SPOCS labels than the `GspSpec` labels, and is on par with the `GspPhot` labels. This is true for both the full sample and the high-SNR subsample. Beyond this, our model is useful for predicting labels that are missing from either or both of the Gaia catalogs (plotted as vertical lines in Figure 5). As shown in the figures, Gaia is missing labels for a number of planet hosts in the sample. This is particularly true for the RVS-derived metallicity and $v_{\rm broad}$ from `GspSpec` (middle panels of rows 3 and 4). Our Cannon model is able to fill these gaps, since it predicts all five labels for every star with an RVS spectrum. Beyond this, our model computes [$\alpha$/Fe], which is missing entirely from both Gaia catalogs.

### 6. Applications for Identifying Anomalous Spectra

In addition to stellar label determination, our model can be used to identify spectra that deviate from the main-sequence single stars that our model was trained on. In this section, we describe metrics that our model computes to flag these anomalous spectra and instructions on how to interpret them. We also provide specific examples of these cases.

Our model computes four metrics that can be used to identify anomalous spectra. The first two, $\chi^2_{\rm Ca}$ and $\Delta\chi^2$, are designed specifically to identify chromospherically active stars and binaries, respectively, and are described in detail in the following sections. The third $\chi^2$ metric measures how well The Cannon model fits the data. This value is minimized during the model's test step (see Section 2.2), and is large for any spectrum that is not well fit by a main-sequence single-star model. The fourth metric is the $\rho(l_n)$ metric described in Section 4.3. A low $\rho(l_n)$ indicates that the model had to extrapolate to increase its flexibility to describe the spectrum. In this case, the fit may or may not have a low $\chi^2$, and the inferred labels are unreliable.

To determine reasonable ranges of our metrics that describe main-sequence stars for which our model can compute accurate labels, we plot the distribution of these metrics in Figure 6 for a sample of main-sequence single stars. This sample comes from Table E3 of EB18, and is composed of stars for which no unresolved spectroscopic binaries were detected in the paper. We removed stars with RUWE>1.4 or Gaia `non_single_star=1` to further ensure the sample contained only single stars. We also required the stars to be the main sequence ($\log g$>4) and removed low SNR targets `rvs_spec_sig_to_noise`<50. The histograms in Figure 6 show the distributions of $\chi^2_{\rm Ca}$, $\Delta\chi^2$, $\chi^2$, and $\rho(l_n)$ for our single-star sample. The remainder of this section outlines specific test cases for which these metrics will be outside the nominal range for main-sequence single stars.

### 6.1. Active Stars

While we mask out the Ca-triplet during our model's test step, these spectral features still contain important information about the star. In particular, stars that are chromospherically active, undergoing accretion, or part of a pair of interacting binaries may exhibit calcium emission that gets superimposed on the Ca-triplet absorption features (Lanzafame et al. 2023). Since our model is assumed to predict flux for inactive stars

---
[9] https://gaia-kepler.fun/

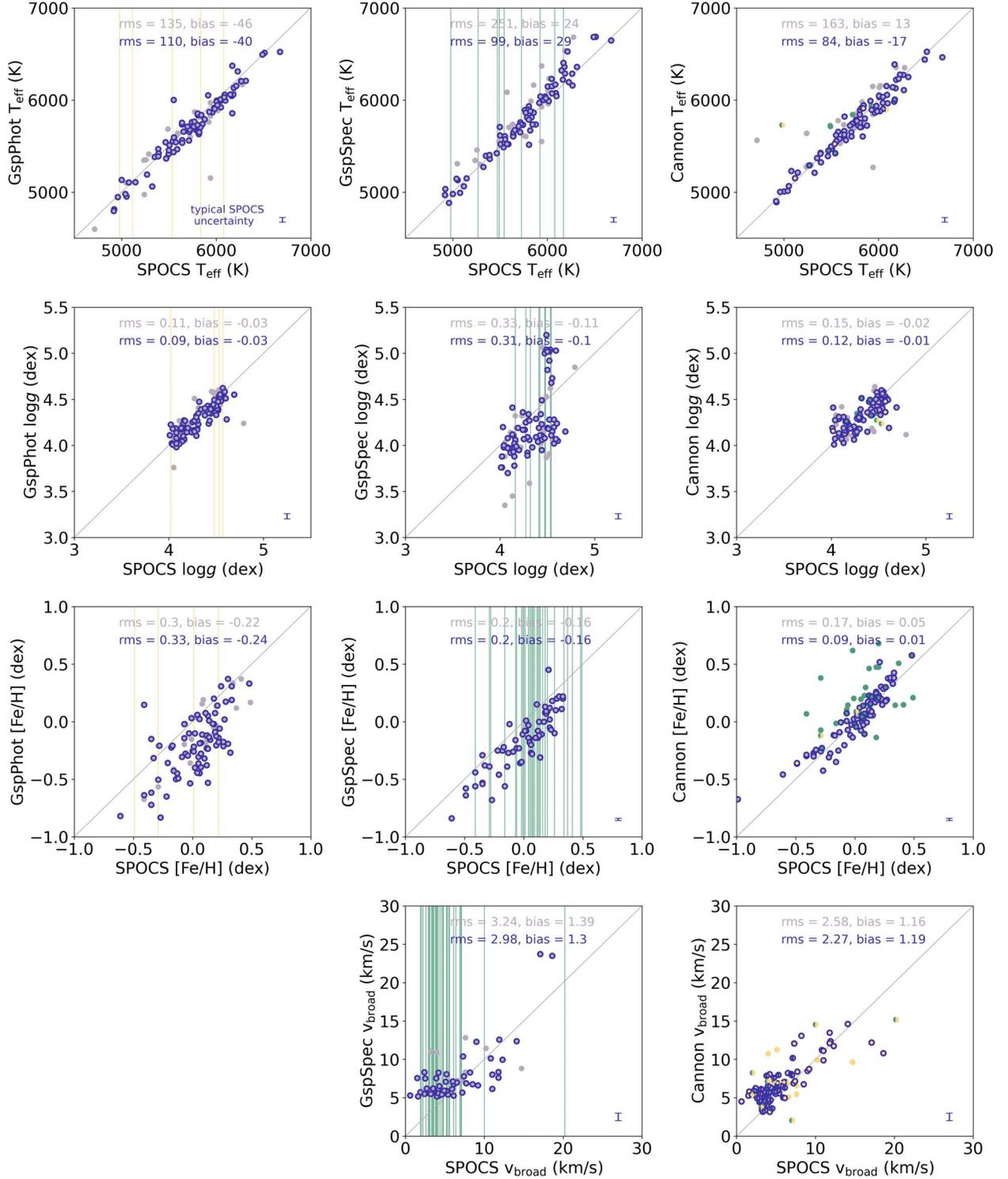

**Figure 5.** Stellar labels from Gaia `GspPhot` (left), Gaia `GspSpec` (middle), and our Cannon model (right) are plotted vs. SPOCS labels for the 107 SPOCS stars with RVS spectra from DR3. The full sample is plotted in gray, and stars with RVS SNR >50 are outlined in blue with the rms scatter and bias in the corresponding colors. Labels missing from Gaia `GspPhot` are represented as yellow lines in the left panel and yellow circles in the right panel. Similarly, labels missing from Gaia `GspSpec` are represented as green lines in the middle panel and green circles in the right panel. Points that are half yellow and half green in the right panel are missing from both Gaia catalogs. We note that (1) we compare [Fe/H] values to [M/H] values for SPOCS and Gaia `GspPhot`, and compute [Fe/H] =fem_gspspec +mh_gspspec for Gaia `GspSpec`, since these catalogs do not report [Fe/H] directly. (2) While Gaia's vbroad is not technically part of the `GspSpec`, we include it because it's derived from the RVS spectrum (Frémat et al. 2023).

(see Section 4.1), activity should manifest as a poor model fit around the calcium features where the model underestimates the flux. To provide a quantitative metric for this activity, we compute the star's $\chi^2_{Ca}$ evaluated only at the wavelengths containing the Ca-triplet—that is, the wavelengths that are masked during the model's test step.

Table 2
Cannon Output Labels for Selection of KIC Stars

| KIC ID | Gaia DR3 ID | $T_{\rm eff}$ (K) | log $g$ (dex) | [Fe/H] (dex) | [$\alpha$/Fe] (dex) | $v_{\rm broad}$ (km s$^{-1}$) | log $\chi^2$ | log $\rho(l_n)$ | log $\chi^2_{\rm Ca}$ | log $\Delta\chi^2$ |
|---|---|---|---|---|---|---|---|---|---|---|
| 1430163 | 2051012738300892416 | 6410 | 4.1 | −0.13 | 0.0 | 12.56 | 3.3 | −2.68 | 2.95 | 0.46 |
| 1434277 | 2051704953890021504 | 5782 | 4.46 | 0.18 | −0.02 | 6.39 | 3.37 | −2.71 | 3.0 | 1.35 |
| 2010835 | 2051082179332118912 | 6061 | 4.35 | −0.23 | 0.07 | 2.89 | 3.29 | −3.55 | 2.83 | 1.08 |
| 2158850 | 2052529106573251584 | 5937 | 4.3 | −0.12 | 0.09 | 7.78 | 3.3 | −3.3 | 2.77 | 1.29 |
| 2285598 | 2099130979202197376 | 5575 | 4.54 | 0.07 | 0.13 | 8.51 | 3.76 | −7.0 | 2.62 | 3.53 |
| 2297349 | 2051098465848130816 | 6149 | 4.2 | −0.47 | 0.14 | 8.09 | 3.38 | −4.12 | 2.79 | 2.21 |
| 2425346 | 2099128848898510336 | 6021 | 4.24 | 0.06 | 0.01 | 7.81 | 3.37 | −2.45 | 2.72 | 1.59 |
| 2441864 | 2052545599247719808 | 5947 | 4.07 | 0.19 | −0.02 | 7.09 | 3.33 | −2.93 | 2.68 | 2.04 |
| 2447773 | 2051844591866815616 | 5930 | 4.11 | 0.15 | −0.01 | 2.28 | 3.38 | −3.21 | 2.69 | 1.56 |
| 2448382 | 2051841598266424448 | 5901 | 4.34 | −0.42 | 0.12 | 4.88 | 3.34 | −3.69 | 2.57 | 1.39 |

**Note.** We report labels and metrics for all main-sequence KIC stars with published RVS spectra in Gaia DR3 with SNR >50, but caution that labels may be unreliable for stars with log $\rho(l_n) < -6$.

(This table is available in its entirety in machine-readable form in the online article.)

We note that this metric is similar to the `activityindex_espcs` that Gaia reports, but with a few notable differences. First, we compute $\chi^2_{\rm Ca}$ by comparing the spectrum to a best-fit Cannon model instead of a template spectrum, thereby measuring the extent to which the target object differs from other stars with otherwise similar labels. This allows us to isolate the effect of chromospheric emission from the natural variance of Ca-triplet flux for stars with different locations on the H-R diagram. Second, our Cannon model allows us to inspect the nature of the star's behavior at the Ca-triplet through the visual inspection of the residuals and the lines' equivalent width, whose sign indicates whether the spectrum exhibits whether the spectrum exhibits calcium emission or absorption relative to the best-fit model. Finally, Gaia does not report the `activityindex_espcs` for all stars. Of the ~1 million stars with RVS spectra, about 1/3 have reported `activityindex_espcs`. Our model can compute reliable $\chi^2_{\rm Ca}$ for any star with an RVS spectrum and reasonable (>50) SNR.

The distribution of $\chi^2_{\rm Ca}$ values for our sample of main-sequence single stars from EB18 is shown in the top right histogram in Figure 6. Objects with $\chi^2_{\rm Ca}$ values that are much larger than the stars in this distribution have Ca-triplet features that differ significantly from those in the training set. An example is shown in the second spectrum from the top for an active dwarf from Lanzafame et al. (2023) with an abnormal emission superimposed in the calcium absorption features. In addition to computing $\chi^2_{\rm Ca}$, our model also enables visual inspection of the star's residuals at the calcium III lines, along with their equivalent widths (which we denote as $W$). Equivalent width values in the single-star sample from EB18 are typically within a range of $0 \pm 0.01$, so $W \approx -0.02$ for this target indicates significant activity-induced calcium emission, along with a large $\chi^2_{\rm Ca}$. However, we caution that a large $\chi^2_{\rm Ca}$ can also be the result of a poor quality of fit at all wavelengths. In this case, the large $\chi^2_{\rm Ca}$ would instead be due to an unresolved binary, evolved star, or noisy spectrum, as shown in the lower panels of the figure. Thus, $\chi^2_{\rm Ca}$ is only a robust activity indicator in the absence of a large $\chi^2$ and small $\rho(l_n)$ that suggests the model fit is unreliable, as described further in Section 6.4.

### 6.2. Unresolved Binaries

Binaries with <1″.8 orbital separations are not resolved by the Gaia RVS instrument and thus produce spectra that are inconsistent with the spectra of single stars (see Cropper et al. 2018, for a description of the RVS instrument design). Among these, binaries with <1 AU physical separations appear as double-lined spectra ("SB2s") and are not processed by Gaia's main spectroscopy pipeline (Collaboration et al. 2023). However, binaries with smaller relative velocities and separations close enough for both stars to fall within the RVS aperture may remain in the current RVS sample as misclassified single stars. These spectral binaries produce two sets of overlapping spectral features and can be identified via Bayesian model comparison of a single-star spectrum with a binary spectrum—that is, a linear combination of two single-star spectra (e.g., EB18). Importantly, for a spectral binary to produce a continuum-normalized spectrum that is significantly different from a single-star spectrum with the labels of the primary, the secondary must contribute significant flux without being too similar to the primary. This implies sensitivity to ~1–200 au binaries with mass ratios anywhere from $q \sim 0.4$ to 0.9.[10]

We model a binary system as combination of a primary star and secondary star with respective labels $l_1$, $l_2$ and radial velocity (RV) offsets $v_1$, $v_2$ relative to the Gaia RVS instrument rest frame.[11] Our Cannon model predicts the rest-frame flux for each of the individual components as a function of stellar labels, $f_1(l_1)$ and $f_2(l_2)$. To account for the RV offsets, we compute wavelength shifts for each binary component relative to the rest frame $\Delta\lambda_{1,2} = \lambda_{1,2} v_{1,2}/c$. We then interpolate The Cannon-predicted fluxes onto wavelength grids shifted by $\Delta\lambda_{1,2}$ to predict the individual spectra of the primary and secondary, $f_1(l_1, v_1)$ and $f_2(l_2, v_2)$. We account for the relative flux contributions of the primary and secondary by multiplying

---

[10] The specific $q$ sensitivity range varies slightly for different instruments, and sensitivity to binaries with $q > 0.9$ can be reached for binaries with larger velocity offsets (see El-Badry et al. 2018a, EB18).

[11] While we expect the relative RV to be small for overlapping spectra, we model nonzero $v_1$ and $v_2$ to account for subpixel RV shifts between the primary and secondary.

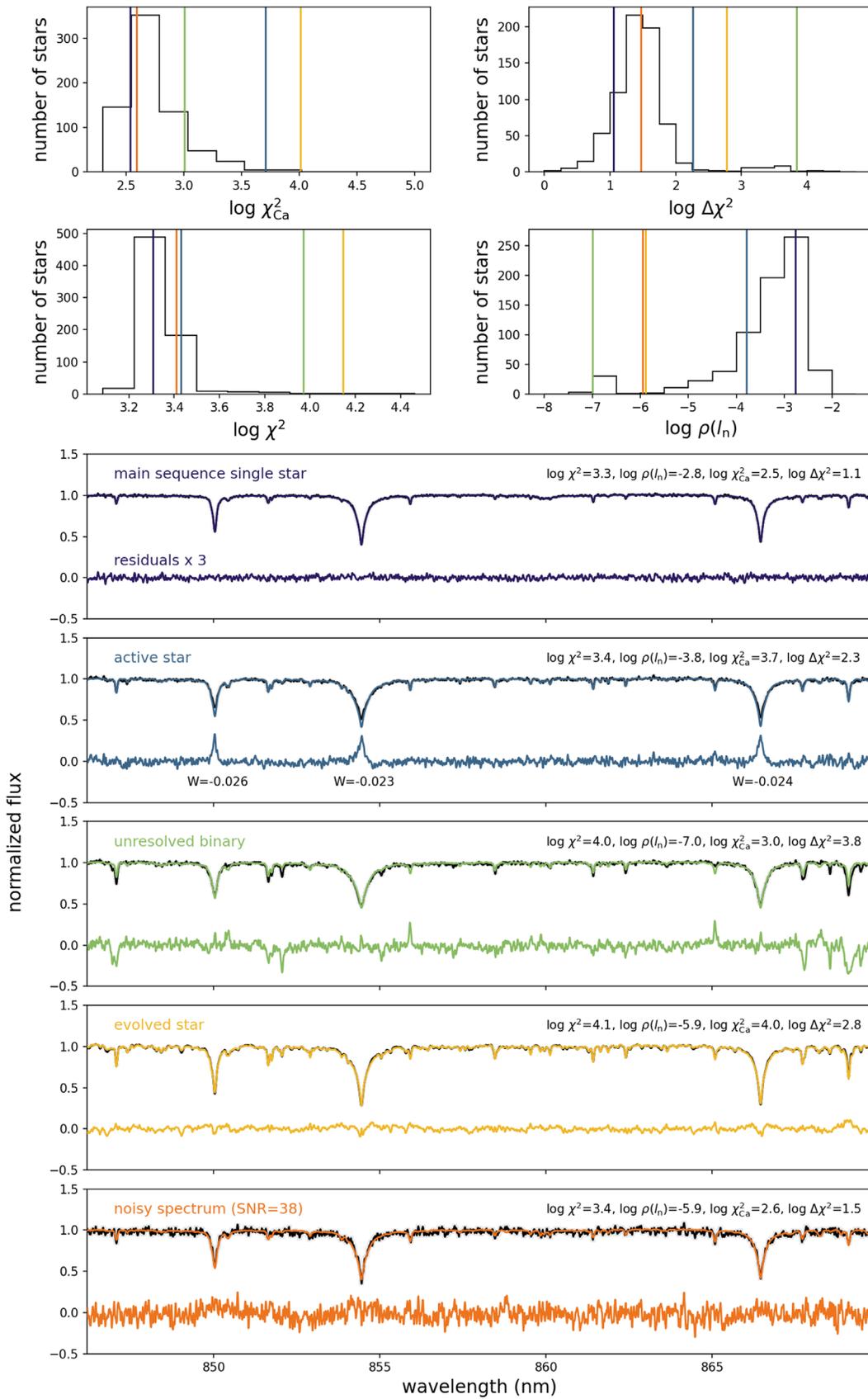

**Figure 6.** Top row: distributions of abnormal star metrics for a vetted sample of main-sequence single stars with SNR>50. Bottom panel: example spectra with best-fit spectra and inferred parameters. Model residuals are plotted below and are inflated by a factor of 3 for visibility. From top to bottom: Gaia DR3 8397285766688669760, a main-sequence single star for comparison; Gaia DR3 4268620287278693120 (HD 176650), an evolved star with log $g \sim 3$; Gaia DR3 3663127407180728704, a noisy (SNR = 30) spectrum of a single star from EB18; Gaia DR3 3174658066485297152, an active dwarf with superimposed calcium emission and absorption from Lanzafame et al. (2023); Gaia DR3 1535964555128078720, an unresolved binary identified in EB18.

$f_1(l_1, v_1)$ and $f_2(l_2, v_2)$ by flux-weighted pre-factors $W_1$ and $W_2$:

$$f_{\text{binary},j} = W_1 \times f_{1,j}(l_1) + W_2 \times f_{2,j}(l_2). \quad (8)$$

where $W_1$ and $W_2$ are calculated based on the flux ratio of the primary and secondary stars, $F_{\text{rel}} = F_1/F_2 = 10^{-(m_1-m_2)/2.5}$. Here, $m_1$ and $m_2$ are the primary and secondary apparent magnitudes, which we compute using empirical magnitude-temperature relations for main-sequence stars reported in Pecaut & Mamajek (2013). $W_1$ and $W_2$ are then derived from $F_{\text{rel}}$ as follows:

$$W_1 = \frac{F_1}{F_1 + F_2} = \frac{F_{\text{rel}}}{F_{\text{rel}} + 1} \quad (9)$$

$$W_2 = 1 - W_1 \quad (10)$$

to assert that the binary flux is a continuum normalized to unity with realistic contributions from the primary and secondary star. The final product is a prediction of the flux at each pixel for a binary system as a function of the primary and secondary labels and RV offsets.

To fit our binary model to an individual RVS spectrum, we repeat the test step described in Section 2.2 but evaluate the likelihood of our binary model instead of the single-star Cannon model. To ensure that our model predicts physically realistic binary systems, we assert that the primary and secondary stars have identical chemical compositions (i.e., [Fe/H] and [α/Fe] are the same for both stars). We also adopt the same training density threshold as our single-star model (Section 4.3) to prevent fitting with unrealistic primary or secondary labels.

One notable difference between the Gaia RVS spectra and the APOGEE spectra used in EB18 is that the Gaia spectra are averaged over anywhere from ∼1 to 100 epochs over a baseline of ∼3 yr. Consequently, for binary systems whose relative radial velocity changes by at least one resolution element $(26\,\text{km s}^{-1})^{12}$ the effect of the secondary is smeared out when the spectra are stacked according to the primary's spectral features and the spectrum does not represent a physical system. In this case, we expect that our model is unreliable, and an inferred RV offset $\geqslant 26\,\text{km s}^{-1}$ should reflect this. For other binaries, the relative RV may not change significantly over the 3 yr baseline, and our model assumption of a constant relative RV between the primary and secondary stars is reasonable.

For a binary to be identified from its RVS spectrum, its spectrum should be significantly better fit by the binary model than the single-star model. This information is contained in the value of $\Delta\chi^2 = \chi^2_{\text{single}} - \chi^2_{\text{binary}}$. It is important to note that due to increased flexibility of the binary model, an RVS spectrum should always be better fit by the binary model than the single-star model. Thus, a robust binary detection requires $\Delta\chi^2$ to be large relative to a sample of single stars. To determine whether this is an effective metric for identifying binaries in the RVS sample, we computed $\Delta\chi^2$ for a sample of known single stars and binaries identified by EB18 (Table E3) with RVS SNR $\geqslant 50$. Our results are shown in Figure 7.

As we can see from this figure, $\Delta\chi^2$ does not highlight a clear separation binaries and single stars in most cases the $\log\Delta\chi^2 = 1$–3 regime encapsulates both 85% of single stars and 87% of binaries. However, we find that a $\log\Delta\chi^2 > 3$ is suggestive of binarity. Indeed, while this region contains stars from both the single-star and binary samples from EB18, the single-star spectra in this region exhibit similar behavior to the binaries, in that the sign of the single-star fit residuals is feature dependent, and the magnitude of the residuals is significantly reduced for the binary fit (see Figure 1 of El-Badry et al. 2018a). Our validation results for abundance labels Section 5.1 indicate that this behavior cannot be attributed to the model inferring incorrect abundances, thus we hypothesize that these stars are true binaries that were missed by EB18. We also find that the $\log\Delta\chi^2 < 1$ region of this figure is largely populated by single stars, with only 1.5% contamination from known binaries. This suggests that our $\Delta\chi^2$ metric can be used in conjunction with other metrics that Gaia reports (RUWE, non_single_star) to rule out the possibility of binarity as opposed to definitively detecting them. Note that $\Delta\chi^2$ refers specifically to spectral binarity, since the system could be a wide binary with a resolved companion and appear as a single star in its spectrum. To summarize, while our $\Delta\chi^2$ metric does not provide a definitive test for binarity for spectra with $\log\Delta\chi^2 = 1$–3, a $\log\Delta\chi^2 > 3$ and $\log\Delta\chi^2 < 1$ are suggestive of binary and single-star scenarios, respectively. We include an example of a binary from EB18 with a $\log\Delta\chi^2 > 3$ in Figure 6.

---
[12] $c/R = 26\,\text{km s}^{-1}$ for the Gaia RVS instrument's $R = 11{,}500$, see Cropper et al. (2018).

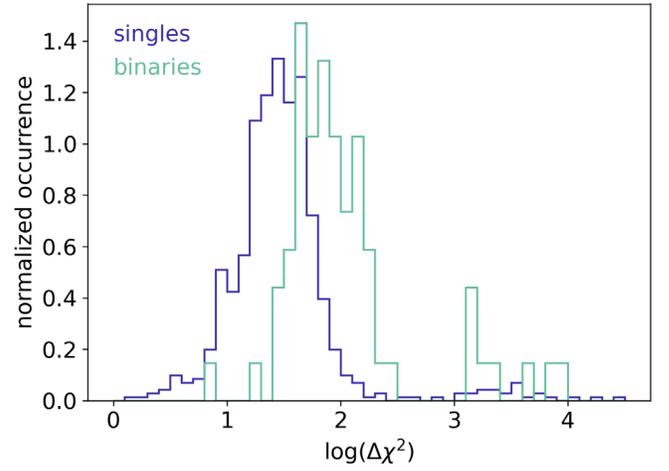

**Figure 7.** Distribution of $\Delta\chi^2$ for known single stars and binaries identified in EB18. Although most stars fall into the $\log\Delta\chi^2 = 1$–3 region where our binary test is inconclusive, a $\log\Delta\chi^2 > 3$ is suggestive of binarity and a $\log\Delta\chi^2 < 1$ can rule out the possibility of a spectral binary.

### 6.3. Evolved Stars

Post-main-sequence stars ($\log g < 4$) will not be well described by our model since it was only trained on stars with $\log g \geqslant 4$. In this case, the $\log\rho(l_n)$ will be low ($\lesssim -6$) since the model will need to extrapolate to explain behavior it was not trained on. However, since we limit the degree to which our model is allowed to extrapolate (Section 4.3), the extrapolation will not be enough for our model to provide a good fit, and the $\chi^2$ value should be large as well.

The fact that the model is unable to provide a good fit in these cases can also manifest in a large $\chi^2_{\text{Ca}}$ and $\Delta\chi^2$ for evolved stars, even if they are not active stars or binaries. We can distinguish evolved stars from active stars because active stars will be well fit by our model everywhere except the calcium lines, and should thus have a nominal $\chi^2$. To this same end, we can distinguish between the evolved and binary cases based on their $\Delta\chi^2$ values and residuals, as demonstrated in Figure [6](). Binaries should possess larger $\Delta\chi^2$ values than evolved stars. Additionally, the residuals of the single-star fit to a binary possess characteristic positive and negative spikes in their residuals, while the evolved star shows residuals that are more smooth and typically negative.

### 6.4. Stars with Unreliable Labels

As discussed in Section [4.3](), a low $\rho(l_n)$ is indicative of extrapolation beyond the training domain during the model's test step, in which case the inferred labels are unreliable. While we penalize models with $\log \rho(l_n) < -7$ (Equation ([7]())), main-sequence single stars typically have $\rho(l_n)$ well above this value. This is shown in the bottom right histogram in Figure [6](), which shows the distribution of $\rho(l_n)$ values for main-sequence single stars from EB18. As we can see, these stars typically have $\log \rho(l_n) \geqslant -5.5$. Thus, a $\log \rho(l_n) \lesssim -6$ is indicative of a star that deviates from a main-sequence single star, and The Cannon output labels are unreliable in these cases.

There are a number of reasons that our model may predict unreliable labels:

1. *Poor data quality.* If the spectrum is too noisy, the model may extrapolate to low $\rho(l_n)$ regions of the training domain to increase flexibility and overfit to the data. This is typically the case for stars with Gaia RVS SNR <50 (see Section [5.2]()).
2. *Unresolved Binaries.* As described in Section [6.2](), the single-star model that is used to infer stellar labels cannot adequately describe the spectrum of an unresolved spectral binary. As shown in El-Badry et al. ([2018a]()), this can bias the inferred parameters of the primary. This can also manifest is a low $\rho(l_n)$, as is the case for the binary shown in Figure [6]().
3. *Evolved Stars.* Because our training set was composed solely of main-sequence stars, accurately modeling an evolved star requires labels beyond the model's training domain. Since the training set contained zero information for these stars, the labels inferred from these spectra are not physical.
4. *True labels are outside of the training domain.* Due to the finite size of our training set, there are bound to be main-sequence single stars with labels that are beyond the training domain that do not satisfy any of the conditions listed above. For example, stars with $T_{\rm eff}$ <4000 K, $T_{\rm eff}$ >7000 K, or particularly large [Fe/H] or [$\alpha$/Fe] abundances may not resemble the stars in our training set. By design, our model is not able to accurately predict labels for these stars.

In general, we caution that labels should not be trusted for any spectrum with a $\log \rho(l_n) < -6$. For RVS spectra with SNR >50, the metrics outlined in this section may also be unreliable since the model performance is compromised. However, for spectra with SNR <50, a low $\rho(l_n)$ is correlated with metrics that can accurately flag binaries and evolved stars. In the binary case, the contamination of the spectral lines due to the secondary contributes to both a low $\rho(l_n)$ and a large $\Delta\chi^2$. In the evolved star case, the object should have a low $\rho(l_n)$ and large $\chi^2$, and possibly a large $\chi^2$ and $\chi^2_{\rm Ca}$ as well. Thus, while we cannot trust the inferred labels in these cases, we can still trust the metrics and flag them as astrophysically anomalous. Note that this is not the case for active stars–$\chi^2_{\rm Ca}$ does not correlate with the $\rho(l_n)$ metric, since the labels are computed based on the fit everywhere except the lines where $\chi^2_{\rm Ca}$ is evaluated. As a result, our model can provide both accurate labels and metrics for active stars.

### 7. Conclusions

In this paper, we train a data-driven model on Gaia RVS spectra of main-sequence stars to predict stellar labels with precisions of 72 K in $T_{\rm eff}$, 0.09 dex in $\log g$, 0.06 dex in [Fe/H], 0.05 dex in [$\alpha$/Fe] and 1.9 km s$^{-1}$ in $v_{\rm broad}$, in agreement with SPOCS labels with roughly 20% and 40% improvement over the Gaia GspPhot and GspSpec catalogs, respectively for stars with SNR >50. We also provide extended data tables with labels inferred by our model for stars from the Kepler Input Catalog (KIC) with SNR>50 with known Gaia DR3 source IDs identified using the Gaia–Kepler cross-match database.[13]

In addition to reporting stellar labels, our model also computes metrics to identify active stars, binaries, evolved stars, and cases where the model's inferred stellar parameters are unreliable. We compute these metrics for a sample of main-sequence single stars from EB18 to determine the nominal ranges of these metric values. Objects with metric values outside of these nominal ranges may be flagged as spectroscopic anomalies. We provide guidelines for interpreting the model metrics for these cases in Section [6]().

We note that while our $\Delta\chi^2$ metric is able to recover some of the binaries identified in EB18, the majority of single stars and binaries in the sample lie in the $\log \Delta\chi^2 = 1-3$ regime where this metric is not conclusive. We interpret the difficulties in identifying binaries in the RVS sample as a product of two factors. First, the low SNR for a large number of stars with published RVS spectra means that flux from the secondary star is likely lost to photon noise, even in cases where the SNR >50. Second, the fact that the RVS spectra are epoch averaged over a 3 yr baseline presents a new obstacle for identifying binaries. In the event that the relative RV of the stars in the binary changes by >26 km s$^{-1}$ over the observing baseline, the secondary features may be smeared out in the averaging step. This implies a complicated selection function that depends on the binary's mass ratio, orbital phase, and separation, and leaves fewer detectable binaries than a similar study with single-epoch spectra. However, it is likely that single-epoch spectra with SNR $\geqslant$100 will be more amenable to identifying binaries via composite Cannon models like the one we develop here. The number of spectra that fit this criterion is large for planet hosts and is expected to grow as Kepler and TESS planet hosts continue to be monitored with instruments like Keck-HIRES and the Keck Planet Finder.

Our model is publicly available at https://github.com/isabelangelo/gaiaspec. The model reports labels and metrics for stars with published Gaia RVS spectra, and provides accompanying plots to visualize and contextualize the best-fit spectra and their corresponding labels and metrics. Our model enables users to classify individual targets of interest, and is particularly useful for TESS planet hosts, many of which have published RVS spectra in Gaia DR3 with reasonable SNR. Beyond this, our code allows these methods to be readily applied to the ~10 million RVS spectra that are expected to be published in future Gaia data releases.

---

[13] https://gaia-kepler.fun/

## Acknowledgments

This work made use of the gaia-kepler.fun cross-match database created by Megan Bedell. I.A. is supported by the Thacher Fellowship at UCLA. I.A. would also like to thank Adrian Price-Whelan, Dan Foreman-Mackey, Kareem El-Badry, Greg Gilbert, and Courtney Carter for their helpful insights on the analysis portion of this paper.

*Software:* astropy (Astropy Collaboration et al. 2013, 2018, 2022), numpy (Van Der Walt et al. 2011), matplotlib (Hunter 2007), scipy (Jones et al. 2001), pandas (McKinney 2010), thecannon (Casey et al. 2016).

## Appendix
## Model Coefficients

As described in Section 2, the trained model coefficients $\theta_j$ describe the model flux dependence on the stellar labels. The magnitude of these coefficient vectors is related to how strongly the model depends on a given label or combination of labels at a particular pixel. During the training step, The Cannon computes coefficient vectors for all linear and quadratic terms in our model, and we show the linear terms corresponding to each label here (Figure 8). We also show the model scatter, which was computed empirically as outlined in Section 4.2.

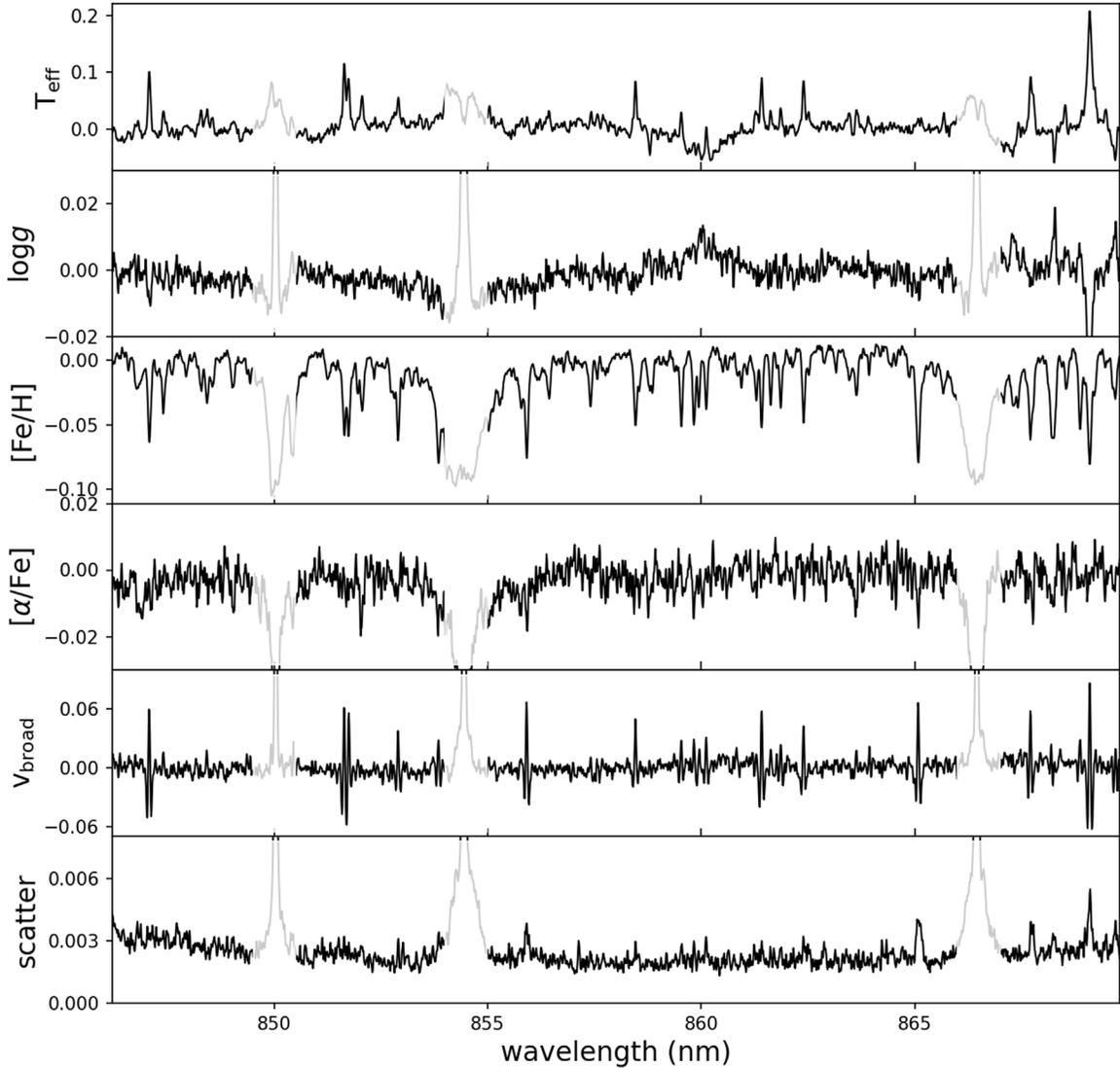

**Figure 8.** First-order coefficient vectors (i.e., spectral derivatives) of our trained model for each of our training labels (top five panels), and model scatter (bottom panel) as a function of the Gaia RVS wavelength scale. For each label, peaks in the coefficient vectors represent pixels in the spectrum that contain the most information for that particular label.

## ORCID iDs

Isabel Angelo  https://orcid.org/0000-0002-9751-2664
Megan Bedell  https://orcid.org/0000-0001-9907-7742
Erik Petigura  https://orcid.org/0000-0003-0967-2893
Melissa Ness  https://orcid.org/0000-0001-5082-6693